# Black-Scholes Model

Mathematical Modeling and Continuous/Discrete Simulation

Francesco Romaggi
S4995443@studenti.unige.it
francesco.romaggi@gmail.com

University of Genoa, Department of
Naval, Electrical, Electronic and Telecommunications Engineering (DITEN)

*Engineering Technology for Strategy and Security*

April 2025

**Abstract**

The main purpose of this article is to give a general overview and understanding of the first widely used option-pricing model, the Black-Scholes model. The history and context are presented, with the usefulness and implications in the economics world.
A brief review of fundamental calculus concepts is introduced to derive and solve the model. The equation is then resolved using both an analytical (variable separation) and a numerical method (finite differences).
Conclusions are drawn in order to understand how Black-Scholes is employed nowadays.
At the end a handy appendix (A) is written with some economics notions to ease the reader's comprehension of the paper; furthermore a second appendix (B) is given with some code scripts, to allow the reader to put in practice some concepts.

# 1 Introduction

The Black-Scholes model is a mathematical model used to shape the dynamics of markets with financial derivative instruments [1].
The main two contributors of the model were the economists: Myron Scholes and Fischer Black.
Robert C. Merton, who first wrote an academic paper on the subject in the 70s, is sometimes also credited as occasionally it is written: Black-Scholes-Merton model. Merton and Scholes received the 1997 Nobel Memorial Prize in Economic Sciences for their work; although ineligible for the prize because of his death in 1995, Black was mentioned as a fundamental contributor.
The formula consists of a parabolic partial differential equation that prices options in the financial market, more specifically it gives a theoretical estimate of the value of European-style options and shows that the asset has a unique price given the risk of the underlying security and its expected return.
The main principle is to hedge the financial derivative by buying and selling the underlying asset in a specific way to eliminate risk. This type of hedging is called "continuously revised delta hedging".
From a mathematical point of view, the Black–Scholes formula has only one parameter that cannot be directly observed in the market: the average future volatility of the underlying security, though it can be found from the price of other options.

As any model it requires some assumptions in order to calibrate properly. [1]
Suppose we consider a call option written on a stock at time $t$. The option expires at time $T > t$ and has a strike price $K$. It is of European style and can be exercised only at expiration date $T$. Further, the underlying asset price and the related market environment denoted by $S_t$ have the following characteristics:

- The risk-free interest rate is constant at $r$.
- The underlying stock price dynamics are described in continuous time by a stochastic differential equation called Geometric Brownian Motion.
- The stock pays no dividends, and there are no stock splits or other corporate actions during the period $[t, T]$.
- Finally, there are no transaction costs and no bid-ask spreads.

It is important to point out that the Black-Scholes equation deals primarily with European-style option, and specifically for the whole paper it is implied that we are always dealing with call options, the put-option-case can be deduced with likewise reasoning.
A further level of complexity arises from the fact that we are applying partial differential calculus in the field of high finance combined with concepts of stochastic calculus as well. For this purpose appendix (A) contains fundamental concepts that may come in help.

[1] For all economics concepts see appendix (A).

# 2 Fundamentals

At first, we need to introduce some mathematical tools that are essential in order to derive, solve and better understand the Black-Scholes model. So the following subsections contain three fundamental mathematical instruments.

## 2.1 Wiener Process, Brownian Motion, Geometric Brownian Motion, Random Walk

In literature there could be a bit of confusion between the following concepts: Wiener Process, Brownian Motion and Geometric Brownian Motion (GBM). [2], [3], [4]
In general, Wiener Process and Brownian Motion are used interchangeably and sometimes we can find papers that refer to Brownian Motion as the Geometric Brownian Motion.
In order to clarify a bit how things are, we can remark that although Wiener Process and Brownian Motion are often used to indicate the same stochastic process, actually they differ slightly due to their origin.
Brownian Motion was first observed in 1827 by botanist Robert Brown to describe the random motion of particles suspended in a medium[2].
Wiener Process is a continuous-time stochastic process often used to model Brownian Motion. A Wiener Process $W_t$ is characterized by four facts:

1. $W_0 = 0$, hence it starts from 0 .
2. $W$ has continuous paths, hence $W_t$ is continuous in $t$.
3. $W$ has independent increments, for every $t > 0$, the future increments $(W_{t+u} - W_t)$, with $u \geq 0$ small, are independents of the past values $W_s, \quad s < t$.
4. $W$ has Gaussian increments, hence $(W_{t+u} - W_t) \sim N(0, u)$.
   Where $N(\mu, \sigma^2)$ is the normal distribution with expected values $\mu$ and variance $\sigma^2$.

For the sake of completeness, it is important to enounce briefly what the Random walk is, since it is often involved when talking about Wiener Process. A Random Walk (sometimes Drunkard's Walk) is a stochastic process that describes a path consisting in a succession of random steps on some mathematical space.
The Wiener process can be seen as the continuous-time limit of a random walk, in which the size of the steps becomes smaller and smaller, as the frequency of steps increases.

[2]By medium we intend liquid or gas, in the specific Brown was looking through a microscope at pollen of the plant Clarkia pulchella immersed in water.

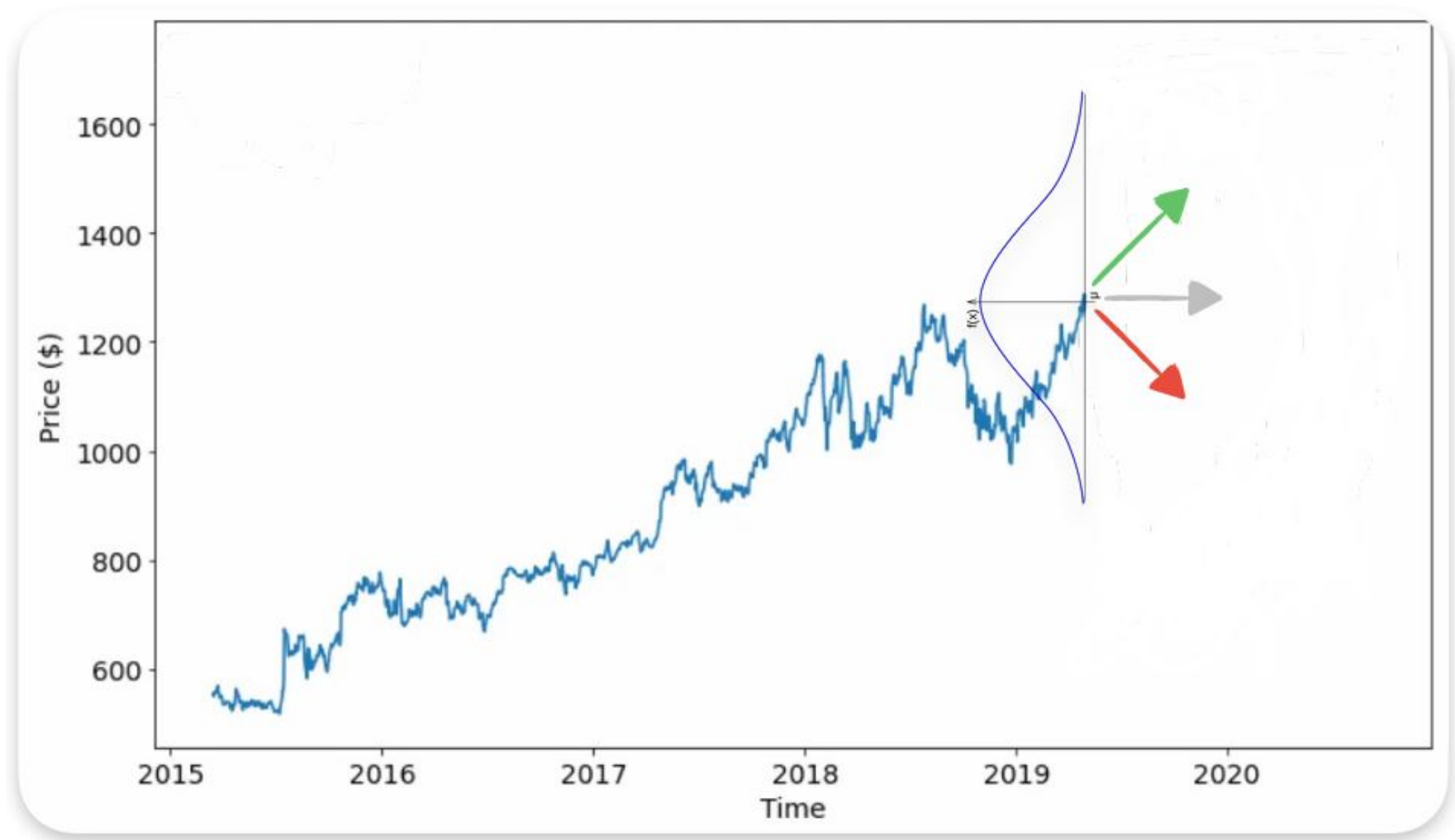

Figure 1: Gaussian increments in the stock market, the future price of a stock is based on a Normal distribution centered on the current value

### 2.1.1 Geometric Brownian Motion

A Geometric Brownian Motion (GBM), is a continuous-time stochastic process in which the logarithm of the randomly varying quantity follows a Wiener Process with drift.
A stochastic process $S_t$ is said to follow a GBM if it satisfies the following stochastic differential equation:

$$dS_t = \mu S_t dt + \sigma S_t dW_t \tag{1}$$

where $W_t$ is a Wiener Process, $\mu$ is the drift, and $\sigma$ is the volatility.
The former parameter is used to model deterministic trends, so expected changes in $S_t$, while the latter parameter models unpredictable events occurring during the motion. These parameters are constants.

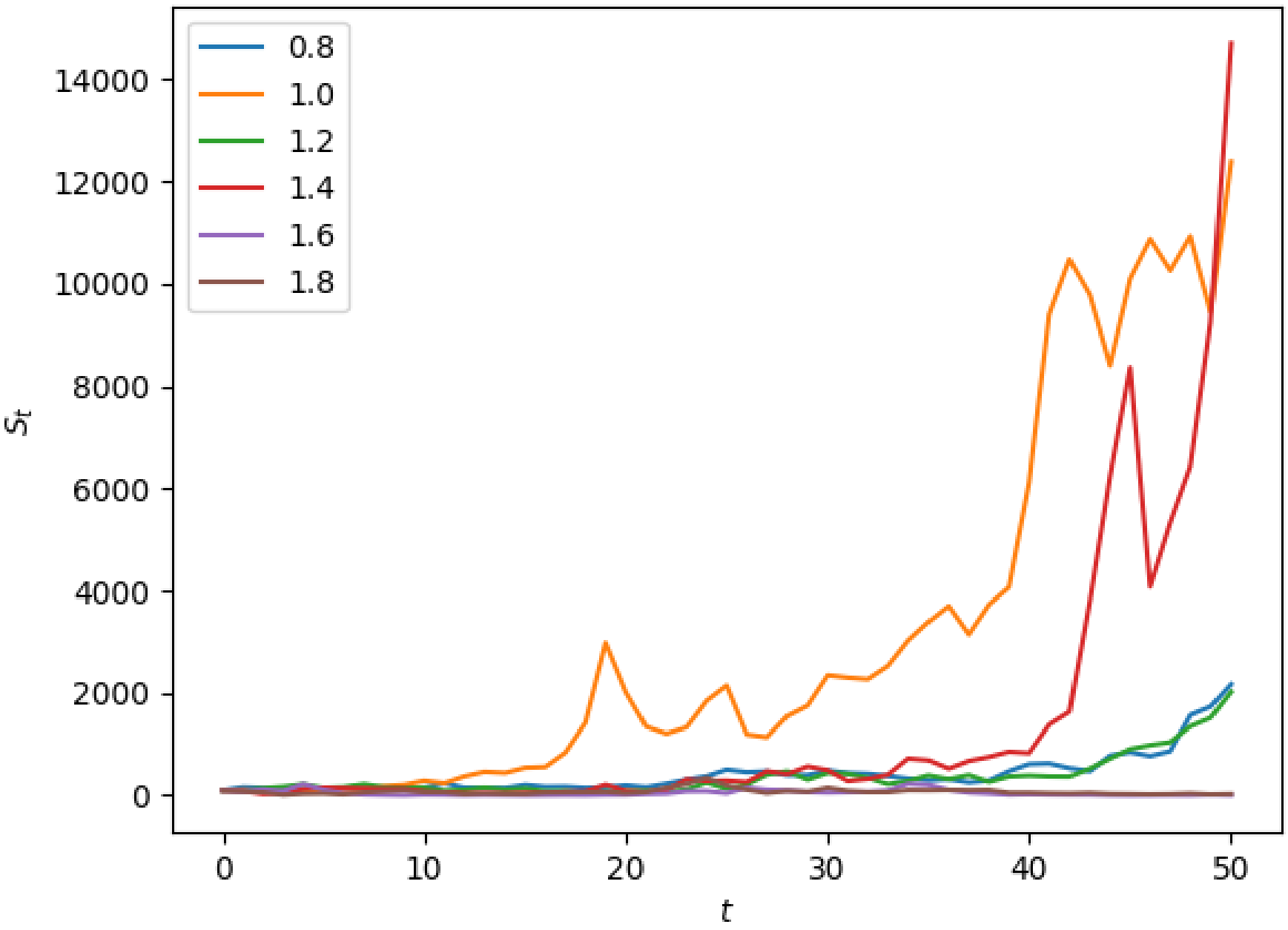

Figure 2: Geometric Brownian Motion with $\mu = 1$ and different $\sigma$, (see appendix (B))

GBM is often used to model financial securities with no specified assumptions.
It shapes the fluctuations of price of the underlying asset in the Black-Scholes model.

## 2.2 Itô's lemma

The core mathematical tool that allows the connection between the GBM and the Black-Scholes equation is given by Ito's lemma.
Ito's lemma is an identity that gives the differential of a time-dependent function of a stochastic process, it can be seen as the stochastic counterpart of the chain rule for derivatives.

Given the stochastic differential equation:

$$dS_t = \mu_t dt + \sigma_t dW_t$$

Where $S_t$ is a stochastic process (or better: a Ito's process), $W_t$ is a Wiener Process, the functions $\mu_t, \sigma_t$ are deterministic functions of time.
We now introduce a function $f$, of class $C2$, then:

- $f(S_t, t)$ is stochastic process, Ito's process
- which leads us to the lemma:

$$df(S_t, t) = \left( \mu \frac{\partial f}{\partial S} + \frac{\partial f}{\partial t} + \frac{1}{2}\sigma^2 \frac{\partial^2 f}{\partial S^2} \right) dt + \sigma \frac{\partial f}{\partial S} dW_t \tag{2}$$

Ito's lemma can be deduced from stochastic calculus and Taylor series.

## 2.3 Second order Cauchy-Euler equation

The first approach we will implement for solving the Black-Scholes equation is analytical via variable separation, and to do so it is necessary to introduce one last fundamental tool: the Cauchy-Euler equation. In mathematics a Cauchy-Euler equation is a linear homogeneous ordinary differential equation with variable coefficients. The most common Cauchy-Euler equation is the second-order, which appears in a number of physics and engineering applications. The second-order equation is:

$$x^2 \frac{d^2y}{dx^2} + ax\frac{dy}{dx} + by = 0 \tag{3}$$

Setting:

$$y = x^m \tag{4}$$

Differentiating a first time:

$$\frac{dy}{dx} = mx^{m-1}$$

Differentiating a second time:

$$\frac{d^2y}{dx^2} = m(m-1)x^{m-2}$$

If we put them back into equation (3):

$$x^2(m(m-1)x^{m-2}) + ax(mx^{m-1}) + b(x^m) = 0$$

Rearranging and factorizing:

$$m^2 + (a-1)m + b = 0$$

So if we solve for $m$:

$$m_1 = \frac{(1-a) + \sqrt{(a-1)^2 - 4b}}{2}$$

$$m_2 = \frac{(1-a) - \sqrt{(a-1)^2 - 4b}}{2}$$

Hence if we put them back into (4) we get the solution for the original equation (3):

$$y = \alpha x^{\frac{(1-a)+\sqrt{(a-1)^2-4b}}{2}} + \beta x^{\frac{(1-a)-\sqrt{(a-1)^2-4b}}{2}} \tag{5}$$

With $\alpha$, $\beta \in \mathbb{R}$

# 3 Black-Scholes-Merton model

The notation used in this section tends to be as close as possible to the one used in the previous parts in order to facilitate connections and comparisons.

## 3.1 Derivation

There are two different approaches on how to derive the Black-Scholes model: through financial reasoning and some economics abstraction, or through pure mathematics. In this paper we stick to a pure mathematical approach.

We consider a financial derivative (i.e.: a financial instrument whose value is derived from the performance of an underlying asset) whose price is denoted by $f(S_t, t)$; $S_t$ is the time-dependent price of the underlying security, that follows a GBM, given by:

$$dS = \mu S dt + \sigma S dW_t \tag{6}$$

Where $W_t$ is a Wiener Process, $\mu$ is the drift of the security, and $\sigma$ the volatility. The dynamics of $S_t$ are assumed to have a constant percentage variance during infinitesimally short intervals.
It is important to point out that $W_t$ represents the only source of uncertainty in the price history of the security.

A portfolio $\pi$ that can eventually offset potential losses can be given by the so called dynamic hedging:

$$\pi = f - \frac{\partial f}{\partial S} S \tag{7}$$

**$\pi$ represents a portfolio made by an option $f$ and a portion of the underlying stock $S$ proportional to the value of $f$ as it varies on $S$.** In other words $\pi$ contains a certain amount of option and a certain amount of the underlying security, in particular the portion of $S$ is proportional to the value of $f$, and since $f$ depends on $S$ this equation offers a way to offset any kind of loss.
This type of hedging is called dynamic hedging, and in an ideal fair market, (also called efficient market), the interest this portfolio returns, should be equal to the Risk Free Rate, i.e.: the rate of return on an investment that has (close to) zero chance of loss, e.g.: U.S. Treasuries.
The underlying concept is that if there is no additional risk, then there are no extra returns.

If we take into account differentials on equation (7) we get:

$$d\pi = df - \frac{\partial f}{\partial S} dS \tag{8}$$

Now we apply Ito's lemma (2) to the GBM of equation (6) inside equation (8):

$$\begin{aligned} d\pi &= \left( \mu S \frac{\partial f}{\partial S} + \frac{\partial f}{\partial t} + \frac{1}{2}\sigma^2 S^2 \frac{\partial^2 f}{\partial S^2} \right) dt + \sigma S \frac{\partial f}{\partial S} dW_t - \mu S \frac{\partial f}{\partial S} dt - \sigma S \frac{\partial f}{\partial S} dW_t = \\ &= \left( \frac{\partial f}{\partial t} + \frac{1}{2}\sigma^2 S^2 \frac{\partial^2 f}{\partial S^2} \right) dt \end{aligned} \tag{9}$$

**The right hand side of equation (9) is free from diffusion term $dW_t$, that is the one related to randomness, hence the portfolio is riskless over an infinitesimal time interval.** Hence we define a constant $r$ that is the Risk Free Rate already explained, that allows us to mathematically write this statement as it follows:

$$d\pi = r\pi dt = r \left( f - \frac{\partial f}{\partial S} \right) S dt \tag{10}$$

Now if we combine (9) and (10) we obtain exactly the Black-Scholes model: [5], [6]

$$rS\frac{\partial f}{\partial S} + \frac{\partial f}{\partial t} + \frac{1}{2}\sigma^2 S^2 \frac{\partial^2 f}{\partial S^2} - rf = 0 \tag{11}$$

That is a parabolic partial differential equation, which brings in relation the price of an option $f$ with the underlying security $S$ in a market without arbitrage.

## 3.2 Solution

In literature exist many methods to solve the Black-Scholes model, they can be gathered into two big categories:

- **Analytical solutions**, when the Black-Scholes model is applied in the most generic cases, it is possible to find a closed-form solution. This can be achieved through different methods which the most remarkable are:
  - using a change of variable to reconduct to the Fourier heat equation;
  - variable separation method;
  - using Fourier or Laplace transform.
- **Numerical solutions**, when the Black-Scholes model is declined in some specific cases, such as particular options or special securities, the only solutions that can be found are based on numerical analysis. This can be achieved through different numerical analysis methods such as:
  - Finite differences;
  - Finite elements;
  - Binomial trees;
  - Monte Carlo simulations;

We choose and develop a method for both categories.
As we have already mentioned before, through this article we deal with call options, the case of put options can be deduced with likewise reasoning.

### 3.2.1 Analytical solution, variable separation

For the sake of brevity and simplicity we use the variable separation method, that is a correct and rigorous method that works under a strong assumption: **the price of the stock $S$ is treated as a generic independent variable, that does not explicitly depend on time.** [7] This is the reason why the solution we will get, differs a bit from the most well known solution.
In their first publications Black and Scholes solved the equation using variable separation too.

Given the equation (11) already seen:

$$rS\frac{\partial f}{\partial S} + \frac{\partial f}{\partial t} + \frac{1}{2}\sigma^2 S^2 \frac{\partial^2 f}{\partial S^2} - rf = 0$$

We apply variable separation to the option variable $f(S,t)$:

$$f(S,t) = A(s)B(t) \tag{12}$$

So the derivatives become:

$$\frac{\partial f}{\partial S} = A'(s)B(t)$$

$$\frac{\partial^2 f}{\partial S^2} = A''(s)B(t)$$

$$\frac{\partial f}{\partial t} = A(s)\dot{B}(t)$$

We also apply a small renaming in order to compact a bit the calculations:

$$a = \frac{1}{2}\sigma^2$$

Now we plug in everything back in the equation (11):

$$aS^2 A''B + rSA'B + A\dot{B} - rAB = 0$$

If we divide by: $AB$ both terms and we rearrange them a bit:

$$aS^2 \frac{A''}{A} + rS\frac{A'}{A} = r - \frac{\dot{B}}{B}$$

So we now are in the case of separation constant from separation of variables.
Hence we introduce the separation constant $c$:

$$aS^2 \frac{A''}{A} + rS\frac{A'}{A} = c = r - \frac{\dot{B}}{B}$$

with $c \in \mathbb{R}$

So we now get a system of ordinary differential equations:

$$\begin{cases} aS^2 \frac{A''}{A} + rS\frac{A'}{A} = c \\ r - \frac{\dot{B}}{B} = c \end{cases} \tag{13}$$

We rearrange the first equation of the system, we reconduct to the Cauchy-Euler equation seen (3):

$$S^2 A'' + \frac{r}{a}SA' - \frac{c}{a}A = 0$$

Set $A = S^m$

Hence:

$$A(S) = \alpha S^{\frac{\left(1-\frac{2r}{\sigma^2}\right)+\sqrt{\left(\frac{2r}{\sigma^2}-1\right)^2+\frac{8c}{\sigma^2}}}{2}} + \beta S^{\frac{\left(1-\frac{2r}{\sigma^2}\right)-\sqrt{\left(\frac{2r}{\sigma^2}-1\right)^2+\frac{8c}{\sigma^2}}}{2}} \tag{14}$$

We rearrange a bit the second equation of system (13):

$$\dot{B} + (r - c)B = 0$$

Whose solution is:

$$B(t) = \gamma e^{(r-c)t} \tag{15}$$

With $\gamma \in \mathbb{R}$

Hence by plugging altogether back in (12):

$$f(S,t) = \left[\alpha S^{\frac{\left(1-\frac{2r}{\sigma^2}\right)+\sqrt{\left(\frac{2r}{\sigma^2}-1\right)^2+\frac{8c}{\sigma^2}}}{2}} + \beta S^{\frac{\left(1-\frac{2r}{\sigma^2}\right)-\sqrt{\left(\frac{2r}{\sigma^2}-1\right)^2+\frac{8c}{\sigma^2}}}{2}}\right] \gamma e^{(r-c)t} \tag{16}$$

In general boundary conditions for the Black-Scholes equation are well known and trivial, they do not bring any further information but are mathematically required in order to determine a unique and physically meaningful solution. Here we implement the conditions in the case of a call option.
Boundary conditions are based on the values that the option will take at the boundaries.

Condition in the upper limit of $t$, hence as $t = T$, where $T$ is the exercise time of the option:

$$f(S,T) = max(S - K, 0)$$

Where $K$ is the strike price.
This means that the equation calculated at the expiry date must coincide with the payoff.

Conditions in the lower and upper limits of $S$:

$S \to 0$:

$$f(0,t) = 0 \quad \forall t \quad \implies A(0) = 0$$

$S \to +\infty$:

$$\lim_{S\to+\infty} f(S,t) \sim \lim_{S\to+\infty} \left(S - Ke^{-r(T-t)}\right)$$
$$\lim_{S\to+\infty} \left[f(S,t) - (S - Ke^{-r(T-t)})\right] = 0$$

The last condition indicates that as the underlying asset price grows very large, the call option's value approaches the underlying asset's price, so in this scenario the option's price value will be almost identical to owning the underlying asset outright.

It is important to point out that the equation is of the second order in $s$ and of first order in $t$, so we only need to set one boundary condition in $t$ and two in $s$.

### 3.2.2 Numerical solution, finite differences

The main idea is to approximate the infinitesimal increments of the differential equation through finite increments, so to transform a differential equation into a finite differences equation. [8]
First we introduce a change of variables, in order to reconduct the Black-Scholes equation to the Fourier heat equation, exactly in the same way as it would have been if we had implemented this method in the previous subsection. [9]

Change of variables:

$$S = Ke^{x}; \quad t = T - \frac{2}{\sigma^2}\tau$$
$$x = ln(\frac{S}{K}); \quad \tau = \frac{\sigma^2}{2}(T - t)$$

And:

$$f(S,t) = Ke^{(\alpha x - \beta\tau)}u(x,\tau)$$
$$u(x,\tau) = \frac{f(S,t)}{Ke^{(\alpha x - \beta\tau)}}$$

So the equation (11), after some calculations becomes:

$$\frac{\partial u}{\partial \tau} = \frac{\partial^2 u}{\partial x^2} \tag{17}$$

Since numerical analysis such as Finite Difference Methods consist of building the solution step by step by iterations, starting from given solution points, in these cases boundary conditions are essentials.

$$u(x,0) = max\left[e^{\frac{k+1}{2}x} - e^{\frac{k-1}{2}x}, 0\right]$$

$$\lim_{x\to-\infty} u(x,\tau) = 0$$

$$\lim_{x\to+\infty} u(x,\tau) = \lim_{x\to+\infty} \frac{S_{max} - Ke^{-r(T-t)}}{Ke^{\alpha x-\beta\tau}} \sim \lim_{x\to+\infty} e^{(1-\alpha)x-\beta\tau}$$

Notice the approximation in the final limit is due to the fact that as $S$ increases a lot, the term $Ke^{-r(T-t)}$ becomes negligible.

Notice that due to the change of variables we have introduced three new quantities:

$$k = \frac{2r}{\sigma^2}$$

$$\alpha = \frac{1-k}{2}$$

$$\beta = -\frac{(k+1)^2}{4}$$

Now we introduce the finite differences equations:

$$\frac{\partial u}{\partial\tau} \approx \frac{u(x,\tau+\Delta\tau) - u(x,\tau)}{\Delta\tau}$$

$$\frac{\partial u}{\partial\tau} \approx \frac{u(x,\tau) - u(x,\tau-\Delta\tau)}{\Delta\tau}$$

$$\frac{\partial u}{\partial\tau} \approx \frac{u(x,\tau+\Delta\tau) - u(x,\tau-\Delta\tau)}{2\Delta\tau}$$

All equations above are equivalent and they rely on Taylor series and incremental ratio. As $\Delta\tau \to 0$ the errors for the first two equations is $(\Delta\tau)$ and for the second one is $(\Delta\tau)^2$.

For the second derivative in $x$ we can use a mix given by:

$$\frac{\partial^2 u}{\partial x^2} \approx \frac{u(x+\Delta x,\tau) - 2u(x,\tau) + u(x-\Delta x,\tau)}{(\Delta x)^2}$$

The error as $\Delta x \to 0$ is in the order of $(\Delta x)^2$.

We need to introduce a bit of formalism in order to get a compact way to write the final recursive equation.
Imagine to divide the grid of the domain of $x$ in many intervals with amplitude $\Delta x$ and the same for $\tau$ with $\Delta\tau$, the function $u(x,\tau)$ will be valued in the points of coordinates: $m\Delta\tau$ and $n\Delta x$, in this way we get an approximated representation of the surface $u(x,\tau)$ that gets better as the intervals $\Delta$ get smaller.
We indicate with $u_n^m$ the value of the function calculated in the point of coordinates: $(n\Delta x, m\Delta\tau)$.

Now we plug the approximations and the formalism back into the original equation (17).
Since we have defined $\frac{\partial u}{\partial\tau}$ in three distinct ways, we can obtain three distinct ways to express the equation, in particular we present two of them.

Renaming: $\delta = \frac{\Delta\tau}{(\Delta x)^2}$

The first implementation is called explicit method, is very easy to compute but it has a strong limitation as $\delta$ exceeds $\frac{1}{2}$, the scheme becomes unstable (the rounding errors increase at each iteration ). So the errors do not increase if: $0 < \delta \leq \frac{1}{2}$

$$u_n^{m+1} = u_n^m(1-2\delta) + \delta(u_{n+1}^m + u_{n-1}^m)$$

The second implementation instead is the implicit method, which is computationally harder, since at each iteration requires the solution of a linear matrix, but on the other hand it does not need any stability check on $\delta$, the equation is defined as:

$$\frac{u_n^{m+1} - u_n^m}{\Delta\tau} = \frac{u_{n+1}^{m+1} - 2u_n^{m+1} + u_{n-1}^{m+1}}{(\Delta x)^2}$$

$$-\delta u_{n+1}^{m+1} + (1+2\delta)u_n^{m+1} - \delta u_{n-1}^{m+1} = u_n^m \tag{18}$$

The last thing we must do is to specify the ranges of the domains.
$x \in (-\infty, +\infty)$, but in order to mathematically treat this problem we restrict $x$ as: $-N\Delta x \leq x \leq N\Delta x$, where $N \in \mathbb{N}$. On the other hand $\tau$ is already restricted due to the way it was defined at first for the change of variables, so: $\tau \in [0, \frac{\delta^2 T}{2}]$, $\Delta\tau = \frac{\sigma^2 T}{2M}$, with $M \in \mathbb{N}$.
Thanks to this, we implement the boundary conditions:

$$u_{-N}^m = u_{-\infty}(-N\Delta x, m\Delta\tau), 0 < m \leq M$$
$$u_N^m = u_{\infty}(N\Delta x, m\Delta\tau), 0 < m \leq M$$
$$u_n^0 = u_0(n\Delta x), 0 \leq n \leq N$$

The equation (18) can also be written in matrix representation $(2N-1) \times (2N-1)$ as following:

$$\begin{bmatrix} 1+2\delta & -\delta & 0 & ... & ... & 0 \\ -\delta & 1+2\delta & -\delta & ... & ... & 0 \\ 0 & -\delta & ... & ... & ... & ... \\ ... & ... & ... & ... & ... & ... \\ 0 & 0 & ... & ... & -\delta & 1+2\delta \end{bmatrix} \times \begin{bmatrix} u_{-N+1}^{m+1} \\ ... \\ u_0^{m+1} \\ ... \\ u_{N-1}^{m+1} \end{bmatrix} = \begin{bmatrix} u_{-N+1}^{m} \\ ... \\ u_0^{m} \\ ... \\ u_{N-1}^{m} \end{bmatrix} + \delta \begin{bmatrix} u_{-N}^{m+1} \\ 0 \\ ... \\ 0 \\ u_N^{m+1} \end{bmatrix} \tag{19}$$

In order to ease a bit the concept of the matrix above, now take $m$ and $n$ varying as:

$$0 < m < M$$
$$0 < n < N$$

Now the matrix (19) can be written as the following system of equations:

$$\begin{cases} (1+2\delta)u_1^{m+1} - \delta u_2^{m+1} = u_1^m + \delta u_0^{m+1} \\ -\delta u_1^{m+1} + (1+2\delta)u_2^{m+1} - \delta u_3^{m+1} = u_2^m \\ -\delta u_2^{m+1} + (1+2\delta)u_3^{m+1} - \delta u_4^{m+1} = u_3^m \\ -\delta u_3^{m+1} + (1+2\delta)u_4^{m+1} - \delta u_5^{m+1} = u_4^m \\ ... \\ ... \\ ... \\ -\delta u_{N-2}^{m+1} + (1+2\delta)u_{N-1}^{m+1} = u_{N-1}^m + \delta u_N^{m+1} \end{cases} \tag{20}$$

The right hand side of the system contains only the known terms, i.e.: the row vector of the previous iteration. In the specific, the known terms of the first and last equations contain also the boundary conditions in $x$.
At each $m$-$th$ iteration we have to solve a system with N-2 equations and N-2 unknown quantities. Those N-2 unknown quantities will be the row vector of the next row, so the known quantities of the following iteration.
Figure 3 shows a visual representation of the values that the matrix above computes at each iteration, i.e.: the rows of this table $M \times N$, that are the values of $u(x, \tau)$.

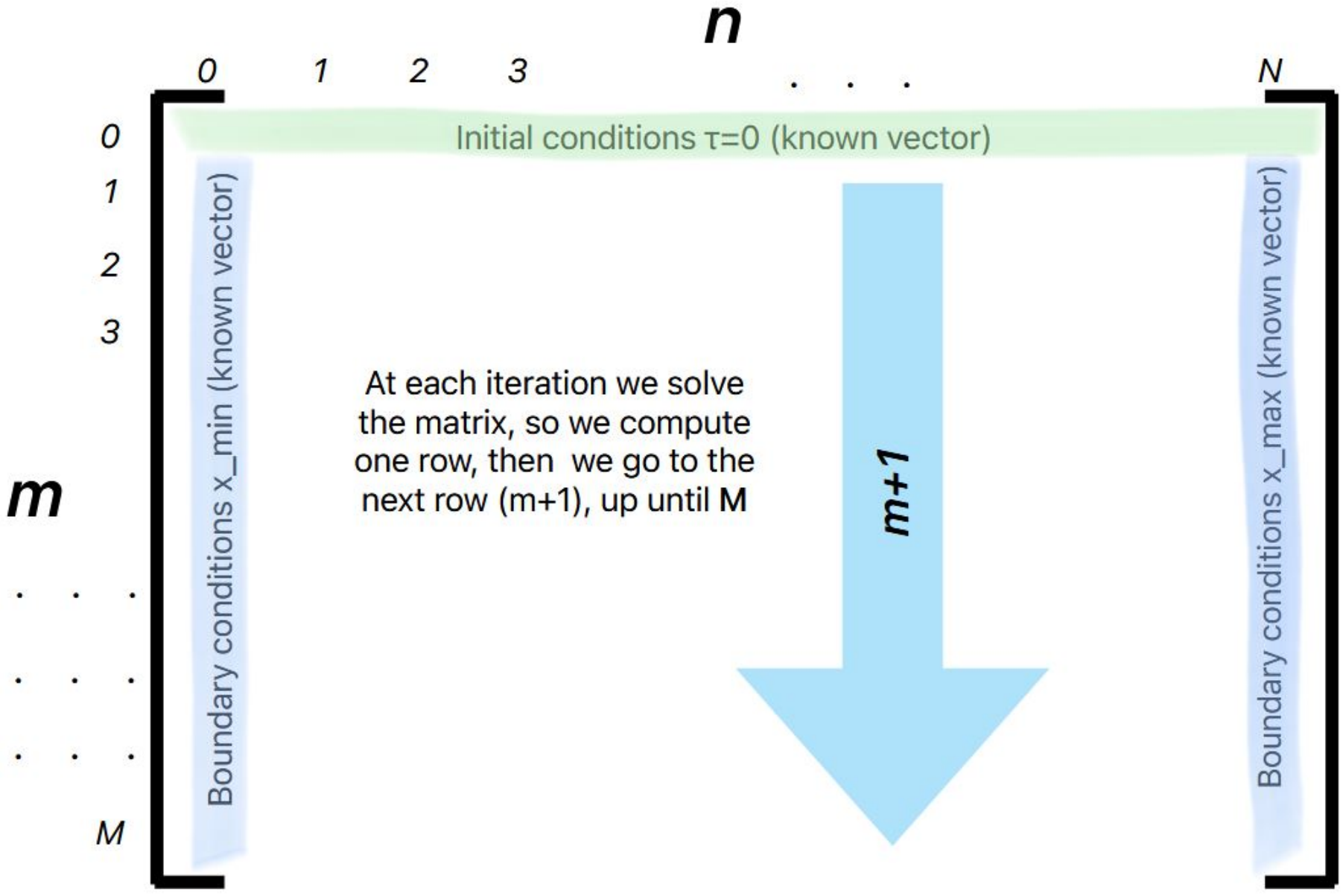

Figure 3: Visual representation of the steps, notice that this is not the matrix of the linear system we are solving, (it is $M \times N$)

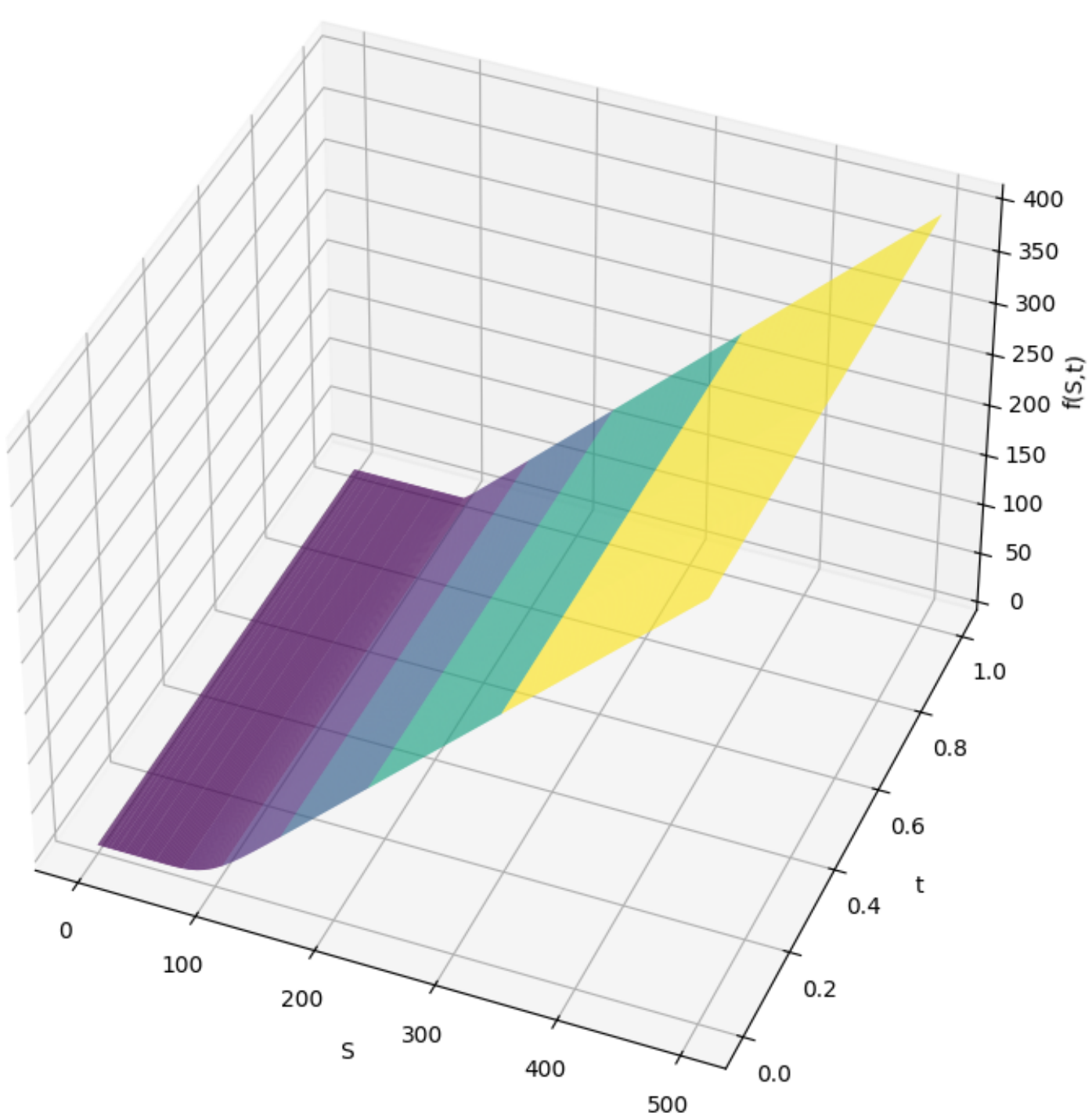

Figure 4: Black-Scholes equation's plot. As S becomes bigger, f shows linear dependence approximating a ramp

The plot in Figure 4 is generated by the Python script implementation for the solution of the Black-Scholes equation with both implicit and explicit method, which can be found in the appendix B. As already discussed, if the $\delta$ parameter is kept $\leq \frac{1}{2}$ the plots for the implicit and explicit methods are exactly the same, otherwise the implicit method would lead to a misleading solution.
The parameters set for Figure 4 are:

- Interest rate ($r$): 5%
- Volatility ($\sigma$): 20%
- Strike price ($K$): 100 of the chosen unit (euro, dollar...)
- Expiration time ($T$): 1 year.
- The maximum price of the underlying security has been chosen arbitrary as five times the strike price (500).

The figure shows the combined behaviors of the security and the call option as the time increases. The ramp takes off as the value of the security overcomes the strike value. In this scenario we expect that the value of the option must increase, because the underlying asset has overcome the value at which we bought it.
Notice that even though from the plot can be quite hard to see, the value of the option changes slightly also over the time horizon, by a tiny diminishing, as time to expiration decreases (time decay).

# 4 Conclusions

The purpose of this paper is to give a general understanding of the Black-Scholes equation, the meaning and reasoning behind it, and different approaches to solve it.
The intention of giving the basic concepts is reflected in the code implementations of the numerical methods adopted, since it has been implemented in a primary straightforward way. Nowadays there many more sophisticated algorithms that are computationally lighter and faster such as the Crank-Nicolson finite differences method or even the Thomas algorithm just for solving tridiagonal matrices.
In any case, on the Internet can be found many different free websites that allow the computation of the Black-Scholes equation given some parameters as input, one can compare the performances of the models implemented in this paper to the ones given by the websites, to gain a general overview of the accuracy.

For the whole paper we dealt with financial derivatives without dwelling on the essence of them, but it is crucial to remark that an option is essentially a volatility instrument. The critical parameter is how much the underlying risk oscillates within a given interval. There are many other risks to manage, The volatility parameter, $\sigma$, may change, interest rates may fluctuate, and option sensitivities may behave unexpectedly. These risks are not "costs" of maintaining the position perhaps, but they affect pricing and play an important role in option trading.
Furthermore, nowadays the financial market is evolving really fast, there are many new technologies and many new financial instruments. In the real world, financial analysts and institutions use ad hoc modifications of the theoretical constructs seen, adapted to make them more realistic.
A specific example of a particular extension of the Black-Scholes formula is the computation of the implied volatility. Instead of calculating the volatility of the underlying asset from historical time series, if we reverse the Black-Scholes formula and we take as input the prices of the European options, we can obtain the so called implied volatility of the underlying asset. This procedure is often used by financial analysts in order to get a sensible approximation of the variances of the securities. From the implied volatility derives the "smile phenomenon" that refers to the observation that the implied volatility of options, when plotted against different strike prices, does not form a flat, constant line as the Black-Scholes model predicts, but instead often resembles a "smile" shape. This can be just a starting point for further digressions and studies that were born as the Black-Scholes model was published.

# References

[1] Principles of Financial Engineering, 3rd Edition. Nov. 2014.
Robert L. Kosowski and Salih N. Neftci.

[2] Wiener Process
ENG https://en.wikipedia.org/wiki/Wiener_process
IT https://it.wikipedia.org/wiki/Processo_di_Wiener

[3] Brownian Motion
ENG https://en.wikipedia.org/wiki/Brownian_motion
IT https://it.wikipedia.org/wiki/Moto_browniano

[4] Geometric Brownian Motion
ENG https://en.wikipedia.org/wiki/Geometric_Brownian_motion
IT https://it.wikipedia.org/wiki/Moto_browniano_geometrico

[5] Black-Scholes Equation Formula
ENG https://en.wikipedia.org/wiki/Black-Scholes_equation
IT https://it.wikipedia.org/wiki/Formula_di_Black_e_Scholes

[6] Black-Scholes Model
ENG https://en.wikipedia.org/wiki/Black-Scholes_model
IT https://it.wikipedia.org/wiki/Modello_di_Black-Scholes-Merton

[7] Solution with variable separation
https://arxiv.org/pdf/1504.03074

[8] Ingegneria Finanziaria, Un'introduzione quantitativa. Mar. 2020.
Barucci, Marsala, Nencini, Sgarra.

[9] Solution with Fourier heat equation
https://digitalcommons.liu.edu/cgi/viewcontent.cgi?article=1074&context=post_honors_theses

[10] YouTube video from Veritasium, for a general and high level overview of the concepts
https://www.youtube.com/watch?v=A5w-dEgIU1M&t=1293s

# A Appendix: economics concepts

In this appendix some key economics concepts that we have come across in the paper are enlisted and explained.

- **Call option**: It is a financial derivative that gives the right but not the obligation to *buy* the underlying security at a later date for a pre-specified price (called strike price). Basically it is bought when the precise is expected to go up.
- **Put option**: It is a financial derivative that gives the right but not the obligation to *sell* the underlying security at a later date for a pre-specified price (called strike price). Basically it is bought when the price is expected to go down.
- **Strike price**: The price at which the option is exercised.
- **European option**: They must be exercised only on the expiry date, whereas *American Option* can be exercised on any date up to expiry.
- **Risk Free Rate**: Is the rate of return of a hypothetical investment with scheduled payments over a fixed period of time that is assumed to meet all payment obligations. Since the risk-free rate can be obtained with no risk, any other investment having some risk will have to have a higher rate of return in order to induce any investors to hold it.
- **Dynamic hedging**: is a risk management strategy where hedge positions are continuously adjusted in response to market changes, aiming to maintain a desired level of risk exposure. It basically means selling/buying and option without taking the opposite side of the trade.
- **Security**: generic term to refer to a general financial asset, such as stock, obligation, treasury bonds...
- **Payoff**: profit or loss realized by the option holder or seller at expiration, based on the difference between the underlying asset's price and the option's strike price.
- **Transaction cost**: expenses incurred beyond the price of a good or service when exchanging it. These costs encompass various aspects of the exchange process, including search, negotiation, and enforcement. In essence, they represent the "friction" or overhead associated with making a deal.
- **Bid-ask spread**: difference between the highest price a buyer is willing to pay for an asset (bid price) and the lowest price a seller is willing to accept (ask price). It represents the dealer's profit margin and a cost of transaction for investors.
- **Arbitrage**: practice involving simultaneously buying and selling an asset in different markets to profit from price differences. This strategy aims to capitalize on temporary discrepancies in pricing by buying an asset where it's cheaper and selling it where it's more expensive.
- **Ideal fair market**: prices fully reflect all available information, meaning there's no way to consistently earn abnormal profits by exploiting market inefficiencies. In an efficient market, prices react to new information quickly and accurately, making it impossible for investors to predict future price movements. In other words the market price is an unbiased estimate of the true value of the investment.

# B Appendix: code

Some relevant code implementations of the concepts seen in sections 2.1.1 and 3.2.2 are shown.

## B.1 Geometric Brownian Motion

```python
# -*- coding: utf-8 -*-
"""
Created on Mon Apr 28 12:45:23 2025

@author: Francesco Romaggi
"""

"""
Notice by playing on the variables mu and sigma the user can generate different
Geometric Brownian Motions

"""

import numpy as np
import matplotlib.pyplot as plt

# 1. PARAMETERS OF THE MODEL
mu = 1 #mean
n = 50 #number of samples
dt = 0.1 #discretization chosen
x0 = 100 #scaling factor
np.random.seed(1) #seed chosen for the random sampling
sigma = np.arange(0.8, 2, 0.2) #array of the sigma values

# 2. MODEL and SIMULATION
x = np.exp(
    (mu - sigma ** 2 / 2) * dt
    + sigma * np.random.normal(0, np.sqrt(dt), size=(len(sigma), n)).T #Wiener part
)
x = np.vstack([np.ones(len(sigma)), x]) #vertical array of ones
x = x0 * x.cumprod(axis=0) #cumulative product for the columns scaled by x0

# 3. PLOTTING
plt.plot(x)
plt.legend(np.round(sigma, 2)) #rounding up to 2 decimals in the legend
plt.xlabel("$t$")
plt.ylabel("$S_t$")
plt.title(
    "Realizations of Geometric Brownian Motion with different variances\n $\mu=1$"
)
plt.show()
```

## B.2 Explicit and implicit finite differences implementations

```python
# -*- coding: utf-8 -*-
"""
Created on Mon Apr 28 12:42:49 2025

@author: Francesco Romaggi
"""

"""
Notice that in numpy np.log stands for the natural logarithm

Notice that sections 1. and 2. of this script are the ones where the user can
interact in order to decline the Black-Scholes equation in the different
scenarios analyzed

Notice the variable: chosen_method is the one that allows the user to choose
the Backward or Forward Finite Difference Method implementation

"""

import numpy as np
import matplotlib.pyplot as plt

#choose which Finite Difference Method you want to iplement ('implicit' or
                                                                    #'explicit')
chosen_method = 'implicit' #otherwise: 'explicit'

# 1. PARAMETERS OF THE MODEL
#choose different parameters based on your needs
r = 0.05 #risk free interest rate
sigma = 0.2 #volatility
K = 100.0 #strike price
S_max = 5 * K #maximum price of the security (we chose aribtrary 5 times K)
T = 1.0 #expiration time (one unit of the chosen time scale)

# 1.1. parameters related to the change of variables
k = 2*r/(sigma**2)
α = (1-k)/2
β = (-(k+1)**2)/4

# 2. DISCRETIZATION
N = 200 #number of [x] steps
M = 2000 #number of [τ] steps

# 3. GRID TIME-SPACE
#space
```

```python
        x_min  = np.log(1e-6) #arbitrary chosen AND ALSO due to how the change of
                              #variables is defined
        x_max  = np.log(S_max/K) #due to how the change of variables is defined
        x = np.linspace(x_min, x_max, N + 1) # #{N+1} points array [x_min; ...; x_max]
        dx = x[1] - x[0]
        #time
        τ_max = (sigma**2)*T/2 #due to how the change of variables is defined
        dτ = τ_max / M #due to how the change of variables is defined
        τ = np.linspace(0, τ_max, M + 1) # #{M+1} points array [0; ...; τ_max]
        #array of the values of the security
        S_grid = K * np.exp(x) #due to how the change of variables is defined

        # 4. INITIALIZATION OF THE GRID OF VALUES
        U = np.zeros((M+1, N+1)) #matrix (M+1)x(N+1) of zeros

        # 6. BOUNDARIES
        #x-->-\infty
        U[:,0]  = 0
        #x-->+\infty
        #U[:,-1] = np.exp((1-a)*x_max - β*τ) <-- this would have been the approximated
                                                                                # version
        #from τ we obtain t_market
        t_market = T - (2 * τ / sigma**2)
        U[:, -1] = (S_max - K * np.exp(-r * (T - t_market))) / (K * np.exp(α*x_max + β*τ))

        #t=0
        U[0,:] = np.maximum(np.exp(((k+1)/2)*x) - np.exp(((k-1)/2)*x), 0)

        # 7.a IMPLICIT SCHEME
        if chosen_method == "implicit":
            # tridiagonal matrix
            delta = dτ / dx**2
            sub = -delta * np.ones(N - 2)
            main = (1 + 2 * delta) * np.ones(N - 1)
            sup = -delta * np.ones(N - 2)
            #construct the tridiagonal matrix based on the diagonals defined above
            A = np.diag(sub, -1) + np.diag(main) + np.diag(sup, 1)

            #implicit scheme implementation
            for j in range(1,M+1):
                #rhs i.e.: right hand side
                rhs = U[j-1, 1:-1].copy() #we take row by row at each iteration
                                          #excluding the boundaries (we leave the first and
                                          #last columns)

                #the following two rows correspond to the contributes of the boundaries
```

```
                rhs[0] += delta * U[j,0] #boundary at: x --> -\infty
                rhs[-1] += delta * U[j,-1] #bounadry at: and x --> +\infty

                #update the solution of the linear system in the matrix, and we'll use
                #it for the next iteration
                U[j, 1:-1] = np.linalg.solve(A, rhs) #we solve row by row each unknown variable

        # 7.b EXPLICIT SCHEME
        if chosen_method == "explicit":
            #definition of the delta
            delta = dτ / dx**2
            assert delta <= 0.5, f"Stability violated (δ={delta:.3f}>0.5) { increase M\
                or decrease dx"
            for i in range(M):
                #we exclude the first and last column
                U[i+1,1:-1] = (
                     delta * U[i,0:-2] #take the whole row excluding the last two elements
                   + (1-2*delta) * U[i,1:-1] #take the row excluding the first and last elements
                   + delta * U[i,2:] #take the whole row starting from the third element
                )

        # 8. TRASFORM BACK TO THE BLACK-SCHOLES MODEL
        #trasformation based on the change of variables
        V = np.zeros_like(U)
        for i in range(M+1):
            V[i,:] = K * np.exp(α*x + β*τ[i]) * U[i,:] #due to how the change of
                                                       #variables is defined

        # 9. PLOTTING SURFACE 3D
        S_mesh, T_mesh = np.meshgrid(S_grid, t_market) #two matrixes composed by the input
                                                       #arrays repeated
        fig = plt.figure() #create the figure object
        ax = fig.add_subplot(111,
                             projection='3d') #adding the 3D subplot to the fig object
        ax.plot_surface(S_mesh, #x axes
                        T_mesh, #y axes
                        V, #z axes
                        rstride=5, #skip 5 rows for drawing (less heavy to draw)
                        cstride=5, #skip 5 columns for drawing (less heavy to draw)
                        cmap='viridis', #heatmap based on the 'viridis scale'
                        alpha=0.9) #opacity
        ax.set_xlabel('S')
        ax.set_ylabel('t')
        ax.set_zlabel('f(S,t)')
```

```
145             ax.set_title('Surface of option price')
146             plt.show()
```